\documentclass[12pt]{iopart}

\usepackage{graphicx}
\usepackage{psfrag}

\def \calU {{\mathcal{U}}}
\def \calW {{\mathcal{W}}}
\def \calX {{\mathcal{X}}}
\def \calY {{\mathcal{Y}}}

\def \calH {{\mathcal{H}}}
\def \calI {{\mathcal{I}}}
\def \calJ {{\mathcal{J}}}
\def \calZ {{\mathcal{Z}}}
\def \calF {{\mathcal{F}}}

\def \bxi{\mbox{\boldmath $\xi$}}
\def \bsxi{\mbox{\footnotesize\boldmath $\xi$}}
\def \bJ {\mbox{\boldmath $J$}}
\def \bs {\mbox{\boldmath $s$}}
\def \bsigma {\mbox{\boldmath $\sigma$}}

\def \sign {\mbox{sign}}
\def \Mup {M_{\mbox{\tt u}}}
\def \Mlo {M_{\mbox{\tt d}}}

\def \argmax {\mathop{\rm argmax}}

\begin{document}

\title{Statistical Mechanics of Broadcast Channels Using Low Density Parity Check Codes}
\author{Kazutaka Nakamura\dag \ , Robert Morelos-Zaragoza\ddag \ ,
David Saad\S \ and Yoshiyuki Kabashima\dag}
\address{\dag \ Department of Computational Intelligence and Systems
Science, Tokyo Institute of Technology, Yokohama 226-8502, Japan}
\address{\ddag \ Advanced Telecommunication Laboratory, SONY Computer Science
Laboratories, Inc., Shinagawa-ku, Tokyo 141-0022, Japan}
\address{\S \ The Neural Computing Research Group, School of Engineering
and Applied Science, Aston University, Birmingham B4 7ET, UK}

\begin{abstract}
  We investigate the use of Gallager's low-density parity-check (LDPC)
  codes in a broadcast channel, one of the fundamental models in
  network information theory. Combining linear codes is a standard
  technique in practical network communication schemes and is known to
  provide better performance than simple timesharing methods when
  algebraic codes are used. The statistical physics based analysis
  shows that the practical performance of the suggested method,
  achieved by employing the belief propagation algorithm, is superior
  to that of LDPC based timesharing codes while the best performance,
  when received transmissions are optimally decoded, is bounded by the
  timesharing limit.
\end{abstract}


\submitto{\JPA}

\maketitle

\section{Introduction}               
Progress in digital communication technologies has dramatically
increased the information flow in both wired and wireless channels.
This makes the role of generic coding techniques, such as
error-correcting codes and data compression, more important. As most
existing codes are constructed for simple point-to-point
communication, they do not necessarily provide optimal performance in
multi-terminal communication such as the inter-net, mobile phones and
satellite communication. Therefore, designing improved codes that
utilize characteristic properties of these media is a promising
direction for enhancing the performance of multi-terminal
communication.

The broadcast channel is a standard multi-terminal communication
channel composed of a single sender and multiple receivers, and is
characteristic of TV and radio broadcasting. Unlike point-to-point
communication, the sender (TV station) simultaneously broadcasts
multiple messages (TV programs) to many receivers (TV sets)
simultaneously via noisy channels. This implies that constructing a
jointly optimal code with respect to the multiple channels may provide
improved performance (i.e., higher capacity) than that of the
time-sharing scheme, whereby separate optimally designed code are used
for each channel. Actually, Cover showed that jointly optimized codes
can have a larger capacity region, where error free communication
becomes possible, than that of timesharing
codes~\cite{Cover_limit,Cover_Book}. However, his proof is
non-constructive and the search for better practical codes for
broadcast channels is still an important topic in information theory.

The purpose of this paper is to devise and analyze an improved
practical code for a broadcast channel by linearly combining
Low-Density Parity-Check (LDPC) codes, which have been shown to
provide nearly optimal performance for single
channels~\cite{Gallager,MN_codes,Davey}. For Reed-Solomon and BCH
codes, which are standard suboptimal codes, it has been reported that
combining codes linearly results in superior performance with respect
to a timeshared transmission~\cite{Algebraic_code,MacWilliams}. This
provides the motivation for the current study, investigating the
performance of linearly combined LDPC codes.

This paper is organized as follows. In the next section, we introduce
the general framework for broadcast channels. Unlike simple
communication channels, the optimal communication performance is still
unknown for most broadcast channels, which would make it difficult to
evaluate the performance of the proposed scheme. Therefore, we focus
here on a simple case, termed a {\em degraded channel}, for which the
capacity region has already been obtained. In section 3, an LDPC code
based construction for degraded channels is introduced, and is
subsequently analyzed in section 4 using methods of statistical
physics. In section 5, the performance of the proposed scheme is
evaluated by solving numerically equations that emerge from the
analysis. The final section is devoted to a summary and conclusion.

\section{Broadcast channel}
\begin{figure}
 \psfrag{SUSIKI_1}{$\footnotesize (\calW_1, \calW_2)$}
 \psfrag{SIKI_1_1}{$\footnotesize \rightarrow \calX$}
 \psfrag{SIKI_2}{$\footnotesize P(\calY_1 | \calX)$}
 \psfrag{SIKI_3}{$\footnotesize P(\calY_2 | \calX)$}
 \psfrag{SUSIKI_4}{$\footnotesize \calY_1 ~\rightarrow~ \hat{\calW_1}$}
 \psfrag{SUSIKI_5}{$\footnotesize \calY_2 ~\rightarrow~ \hat{\calW_2}$}
 \psfrag{SUSIKI_6}{$\footnotesize P(\calY_1, \calY_2 | \calX)$}
 \psfrag{R_1}{$\footnotesize R_1$}
 \psfrag{R_2}{$\footnotesize R_2$}
 \psfrag{H_1}{$\footnotesize 1-H(p_1)$}
 \psfrag{H_2}{$\footnotesize 1-H(p_2)$}
  \includegraphics[width=7.5cm,clip]{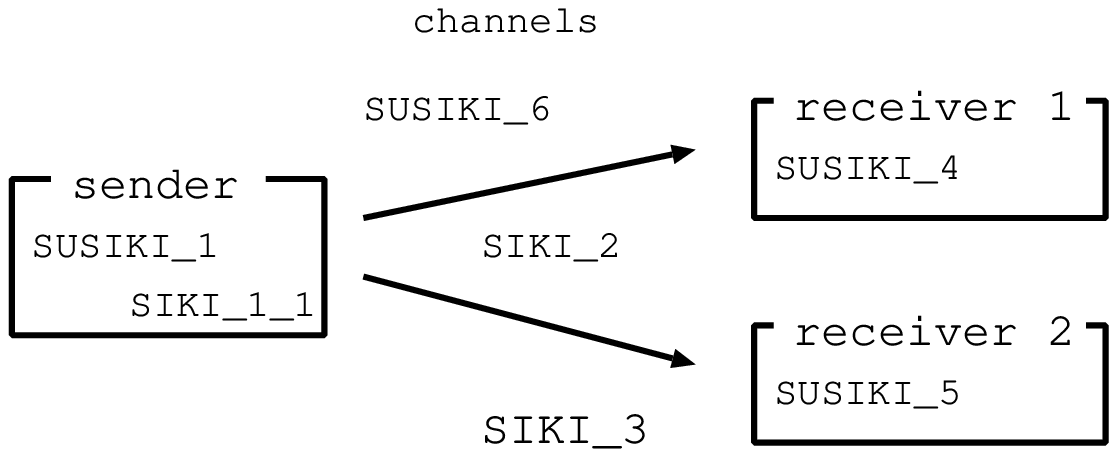}
  \includegraphics[width=7cm,clip]{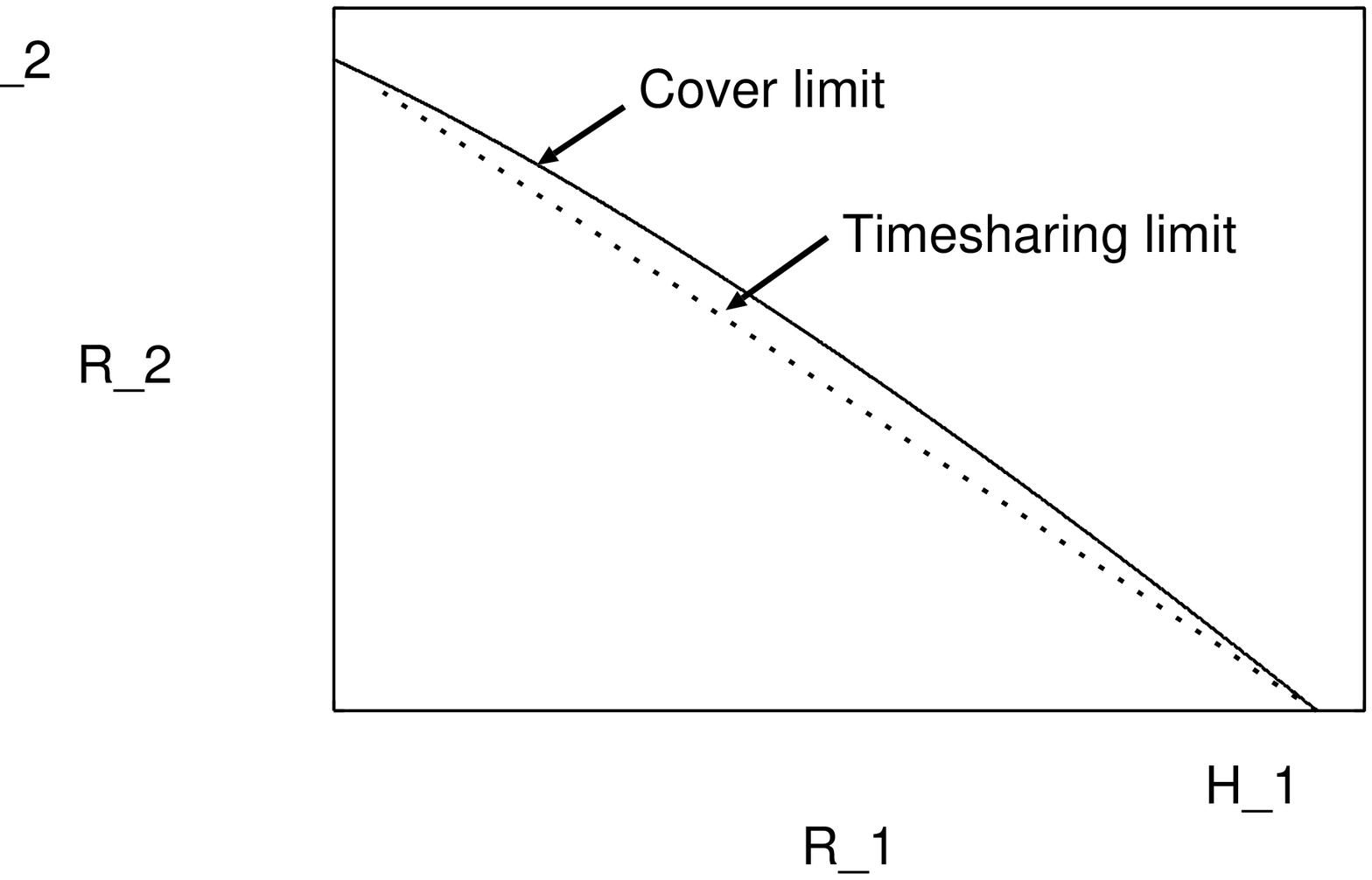}
 \caption{(a): A single sender and two receivers broadcast channel. 
(b): The capacity region in the case of binary symmetric channels. 
The solid curve and the dotted line denote Cover's and timesharing 
limits, respectively. 
}
\label{fig:broadcast}
\end{figure}

In the general framework of broadcast channels,
a single sender broadcasts a codeword composed of 
different messages to multiple receivers. For simplicity, 
we here restrict our attention to the case 
of a single sender and two receivers (Fig.~\ref{fig:broadcast}), 
where one codeword $\calX$ ($N$ bits), comprising two messages
$\calW_1$($R_1 N$ bits) and $\calW_2$($R_2 N$ bits), is sent to two
receivers. As each channel is noisy, receivers $1$ and $2$ obtain two
corrupted codewords $\calY_1$ and $\calY_2$, respectively; this is
modeled by a conditional probability $P(\calY_1, \calY_2|\calX)$. The
received corrupted codewords $\calY_1$ and $\calY_2$ are decoded by
the respective receivers to retrieve only the message addressed to
each of them.

Analogously to the case of single channels, error free communication
becomes possible if the corresponding code rate vector $(R_1,R_2)$
lies within a certain convex region, termed the {\em capacity region},
determined for a given broadcast channel $P(\calY_1, \calY_2|\calX)$
using an infinite code length
$N$~\cite{Cover_Book,BERGMAN}. Evaluation of the capacity region is
one of the fundamental problems in information theory; the problem is
generally difficult and has not yet been solved in general except for
a few special cases.

A broadcast channel $P(\calY_1, \calY_2|\calX)$ is termed {\it
  degraded} if there exists a distribution $P^\prime(\calY_2
|\calY_1)$ such that
\begin{equation}
P(\calY_2 | \calX) = \sum_{\{\calY_1\}} P(\calY_1 |\calX) ~
P^\prime(\calY_2|\calY_1) \ .
\label{eq:degraded_channel}
\end{equation}
The degraded channel is exceptional in the sense that its capacity
region can be analytically obtained as the convex hull of the closure
of all points $(R_1, R_2)$ that satisfy
\begin{equation}
\left\{
 \begin{array}{l}
  R_2 < I(\calU | \calY_2) \\
  R_1 < I(\calX ; \calY_1 | \calU)
   \label{eq:capacity}
 \end{array}
 \right.
\end{equation}
for a certain joint distribution $P(\calU) P(\calX|\calU) P(\calY_1,
\calY_2|\calX)$; where the auxiliary random variable $\calU$ has a
cardinality bounded by $|\calU| \le \min\{|\calX|,|\calY_1|,
|\calY_2|\}$. This region is often called Cover's
capacity~\cite{Cover_limit, Cover_Book} region. Unfortunately, the
derivation of Cover's capacity is non-constructive and offers little
clue to design efficient practical codes. Thus, practical codes for
the degraded broadcast channel has been actively investigated in the
network information theory~\cite{Cover_comment}.

In the case of binary symmetric channels characterized by flip
probabilities $p_1$ and $p_2$, condition~(\ref{eq:degraded_channel})
reduces to an inequality $p_2 > p_1$. Then, the expression of Cover's
capacity is simplified to
\begin{equation}
\left\{
\begin{array}{l}
R_2 < 1 - H_2(\delta * p_2) \\
R_1 < H_2(\delta * p_1) - H(p_1)
\end{array}
\right.
\end{equation}
where a parameter $0<\delta <1$ specifies the optimal 
ratio between $R_1$ and $R_2$; 
$\delta * p = \delta (1-p) + (1-\delta) p$ and $H_2(p)$ 
is Shannon's entropy $ H_2(p) = - p \log_2 p - (1-p) \log_2 (1-p)$.

The solid convex curve in Fig.\ref{fig:broadcast}(b) shows Cover's
limit, i.e., the boundary of Cover's capacity for the binary symmetric
channels. The straight broken line corresponds to the timesharing
capacity, i.e., the achievable capacity by concatenating two
independent codeswords optimized for each channel separately. This is
realized by using $N(1-\alpha)$ and $N \alpha$ bits of
codeword $\calX$ for encoding messages $\calW_1$ and $\calW_2$,
respectively. Here, $0< \alpha <1$ is the code length ratio between
the two messages. This simple concatenation and the limit achievable
by this scheme are often termed the {\em timesharing} and the {\em
  timesharing limit}, respectively. The difference between Cover's
and the timesharing limits indicates the capacity gain obtained by
optimizing a code for the complete broadcasting system in comparison
with respect to optimizing each of the channels separately.

We have to emphasize that achieving the timesharing limit {\em in
practice} is never trivial as there is no known practical code that
saturates Shannon's limit even for a single channel. Therefore, the
design of improved practical codes for broadcasting, by combining
existing codes, devised for single channels, is an important research
topic in coding theory~\cite{MacWilliams}.

\section{Linearly combined codes}
Linearly combined codes is a well-known strategy for designing high
performance communication schemes for broadcast channels using
multiple linear Error-Correcting Codes
(ECC)~\cite{Algebraic_code,MacWilliams}. In this scheme, the first $N
(1-\alpha)$ bits of a codeword are obtained by linearly mixing two
messages $\calW_1$ and $\calW_2$ while the other $N \alpha$ bits are
generated only from $\calW_2$ by some linear transformation. In both
coding and decoding, all operations are typically carried out in
modulo $2$. This method has been developed for algebraic codes, such
as Reed-Solomon and BCH, which are standard codes designed for
relatively short code lengths. For these codes, it is reported that
the minimum distance between codewords is larger than that achieved in
the timesharing scheme, which implies higher robustness against
channel noise~\cite{Algebraic_code,MacWilliams}.

However, it is unclear whether a similar construction also offers 
better performance when different code types are used. Furthermore,
it is theoretically interesting and important to examine whether a
linearly combined code can saturates Cover's limit for infinite code
length ($N$) or not.

Motivated by these questions, we investigate here the ability and
limitations of linearly combined LDPC codes in the limit $N \to
\infty$.

An LDPC code is characterized by a parity check matrix. To
devise a linearly combined coding scheme for LDPC codes, we define a
parity check matrix in an upper triangular form
\begin{eqnarray}
 A = \left(
      \begin{array}{@{\,}cc@{\,}}
       A_1 & A_2 \\
        0 & A_3
        \end{array}
     \right)
\end{eqnarray}
where the sizes of the sub-matrices $A_1, A_2, A_3$ are 
$\left [(1-\alpha) N - R_1 N \right ]\times (1-\alpha) N$,
$\left [(1-\alpha) N - R_1 N \right ]\times \alpha N$, 
$\left [\alpha N - R_2 N \right ]\times \alpha N$, respectively. 
Further, we assume that $A_1, A_2, A_3$ have $K_1, K_2, K_3$ and 
$C_1, C_2, C_3$ non-zero elements per row and column, respectively.
Based on the parity check matrix, 
the generator matrix $G^T$ is constructed as 
\begin{equation}
 G^T = \left(
      \begin{array}{@{\,}cc@{\,}}
       G_1^T & G_2^T \\
        0 & G_3^T
        \end{array}
     \right)
\end{equation}
where $G^T_i~~(i=1,3)$ are constructed systematically
to satisfy the constraints $A_i G_i^T = 0~(\bmod{2})$ and
$G_2^T$ is defined as $ -A_1^T[A_1 A_1^T]^{-1}[A_2 G_3^T]$.
The sizes of these matrices are $(1-\alpha) N \times R_1 N$, 
$\alpha N \times R_2 N$ and $\alpha N \times R_2 N$, respectively.

The sender produces a codeword $\calX$ by taking a product of the
generator matrix $G^T$ and the original messages $(\calW_1,
\calW_2)^T$. Receiving a possibly corrupted codeword, each receiver
evaluates the syndrome vectors $\bJ_i = A \calY_i~(i=1,2)$, which
yield the parity-check equation $\bJ_i = A \bxi_i$. The message
vector $\bxi_i$ can be thought of as having two separate segments
denoted by {\tt u} (up) and {\tt d} (down) later on. The parameter
$\alpha$ controls the error correction ability for the second message;
the transmitted information redundancy increases with $\alpha$
($\alpha > R_2/(R_1+R_2)$). The decoding problem for each
receiver is to find the most probable messages, $\bs_i$ and
$\bsigma_i$, such that the parity check equation
\begin{equation}
  \bJ_i = \left(
         \begin{array}{@{\,}cc@{\,}}
          A_1 & A_2 \\
          0 & A_3
         \end{array}
       \right)
  \left(
   \begin{array}{@{\,}c@{\,}}
    \bs_i \\
    \bsigma_i
   \end{array}
 \right) \quad (\mbox{$i=1,2$})
\label{eq:parity_check}
\end{equation}
is obeyed, and using prior knowledge about the two noise vectors
characterized by the two different channels.

The second receiver has to estimate only the lower part 
of noise vector $\bxi^{\mbox{\tt d}}$, which can be carried out 
using only the lower part of Eq.(\ref{eq:parity_check}).
However, we assume here that both receivers independently solve 
Eq.(\ref{eq:parity_check}) using prior knowledge on their own channels
since one can show that solving the whole equation provides 
the optimal estimation performance for both receivers. 
As Eq.(\ref{eq:parity_check}) has the same form 
for receivers $1$ and $2$, we hereafter omit the subscript $i=1,2$. 

For bit-wise minimization of the error probability
the optimal estimation is given by maximizing the posterior marginal (MPM) 
\begin{equation}
 \hat{\xi}^{\mbox{\tt u}}_i = \arg \max_{s_i \in \{0/1\}} P(s_i |\bJ),~~ 
 \hat{\xi}^{\mbox{\tt d}}_j = \argmax_{\sigma_i \in \{0/1\}} P(\sigma_j |\bJ).
\label{eq:MPM}
\end{equation}
An exact evaluation of Eq.(\ref{eq:MPM}) is generally hard; therefore,
the belief propagation (BP) approximation scheme is widely used as a
practical decoding algorithm. The latter has been shown to be
identical to the Thouless-Anderson-Palmer (TAP) approach in the
current case~\cite{BPvsTAP}.

\section{Statistical mechanics}

\subsection{Macroscopic analysis -- performance evaluation }
In order to evaluate the typical error-correction ability of these
codes in the limit $N \to \infty$, we investigate the behavior of the
MPM decoder using the established methods of statistical mechanics.
We first map the current system to an Ising spin model with finite
connectivity, by employing the binary representation $\{+1,
-1,\times\}$ for the alphabet and operator instead of the
Boolean one $\{0,1,+\}$. This implies that the posterior probability
$P(\bs, \bsigma |\bJ)$ can be expressed as a Boltzmann distribution at
the inverse temperature $\beta=1$ using a Hamiltonian
\begin{eqnarray}
\fl \calH(\bs, \bsigma | \bJ) &=& \lim_{\gamma \to \infty} \left\{
 \gamma \sum_{\{ \calI(K_1), \calJ(K_2) \} } 
   D_{ \calI(K_1), \calJ(K_2)}^{1,2} 
\delta( -J^{\mbox{\tt u}}_{\calI(K_1),\calJ (K_2)} ;
 \prod_{i \in \calI(K_1)} s_i 
 \prod_{j\in \calJ (K_2)} \sigma_j ) \right. \nonumber \\
\fl &{}&\left. + \gamma \sum_{\{ \calJ(K_3)\}} 
 D_{\calJ(K_3)}^3 
 \delta( -J^{\mbox{\tt d}}_{\calJ(K_3)};
 \prod_{j \in \calJ(K_3)} \sigma_j ) \right\}
 - F \sum_{i=1}^{(1-\alpha)N} s_i - F \sum_{j=1}^{\alpha N} \sigma_j,
\label{eq:Hamiltonian}
\end{eqnarray}
where $\calI(K)=\langle i_1,i_2 \cdots, i_{K} \rangle$ denotes the
combination of the $K$ subscripts chosen from the $i=1,2, \cdots,
(1-\alpha)N$ possibilities without duplication (the order is ignored),
and $\calJ(K)=\langle j_1,j_2 \cdots, j_{K} \rangle$ is the $K$
combination from $j=1,2, \cdots, \alpha N$ chosen similarly. The
tensor $D_{\calI(K_1),\calJ(K_2)}^{1,2}$ becomes 1 when its subscripts
agree with the positions of non-zero elements in the parity-check
matrices $A_1$ and $A_2$, and 0 otherwise. The tensor
$D_{\calJ(K_3)}^3$ similarly corresponds to $A_3$. The first and
second terms in Hamiltonian~(\ref{eq:Hamiltonian}) correspond to
Eq.(\ref{eq:parity_check}) while the third and fourth terms are
provided by the prior distribution of the noise. The field $F$
represents the channel noise level; it is set to $\frac{1}{2}\ln
(1-p_1)/p_1$ and $\frac{1}{2}\ln (1-p_2)/p_2$ for the first and the
second receivers, respectively.

In order to simplify the calculation, we first 
employ the gauge transformation 
$s_i ~\to~ s_i \xi_i^{\mbox{\tt u}}, \sigma_j ~\to~ \sigma_j
\xi_j^{\mbox{\tt d}}$, $J^{\rm u}_{\cdots} \to 1$ and
$J^{\rm d}_{\cdots} \to 1$, which 
reduces complicated couplings expressed in the first 
and second terms in Hamiltonian~(\ref{eq:Hamiltonian})
to simple ferromagnetic interactions. 

As the parity check matrices and noise vectors are generated
randomly, we have to perform averages over these 
variables for extracting typical properties of the code.
This can be carried out by the replica method 

$- \beta \calF = \langle \ln \calZ \rangle_{A,\bsxi^{\tt u},\bsxi^{\tt
    d}} = \lim_{n \to 0}(1/n)\ln \langle \calZ^n -1
\rangle_{A,\bsxi^{\tt u},\bsxi^{\tt d}}$, where $Z$ is the partition
function and $\langle \cdots \rangle_{A,\bsxi^{\tt u},\bsxi^{\tt d}}$
represents an average over the parity check matrix $A$ and the noise
vectors $\bxi^{\tt u}$ and $\bxi^{\tt d}$ (i.e., the quenched
variables). This gives rise to three sets of order parameters
\begin{eqnarray}
q_{\{a_1, a_2, \cdots, a_m\}} &= \frac{1}{N}
\sum_{i=1}^{(1-\alpha)N} X_i s_i^{a_1} \dots s_i^{a_m}, \nonumber \\
r_{\{a_1, a_2, \cdots, a_m\}} &= \frac{1}{N}
\sum_{j=1}^{\alpha N} Y_j \sigma_j^{a_1} \dots \sigma_j^{a_m}, \nonumber \\
t_{\{a_1, a_2, \cdots, a_m\}} &= \frac{1}{N}
\sum_{j=1}^{\alpha N} Z_j \sigma_j^{a_1} \dots \sigma_j^{a_m} 
\end{eqnarray}
where $a_1, a_2, \cdots, a_m$ denote the replica indices running from
$1$ to $n$, and their conjugates
$\hat{q}_{\{a_1, a_2, \cdots, a_m\}}$, 
$\hat{r}_{\{a_1, a_2, \cdots, a_m\}}$, 
$\hat{t}_{\{a_1, a_2, \cdots, a_m\}}$. 
The variables $Z_j$ are introduced to express the constraint
of the parity-check matrix $A_3$ as 
\begin{equation}
\delta \left( 
\sum_{\calJ(K_3)/j} D^3_{\calJ(K_3)} - C_3
\right)=\oint \frac{dZ_j}{2\pi} Z_j^{\sum_{\calJ(K_3)/j} 
D^3_{\calJ(K_3)} - (C_3+1)}. 
\end{equation}
The variables $X_i$ and  $Y_j$ are similarly introduced for $A_2$ and $A_3$.

In order to proceed further, one has to make an assumption 
about the symmetry of replica indices. Here we employ the 
simplest replica symmetric (RS) ansatz, expressed in the current case by
$q_{\{a_1, \dots, a_m\}} =   q_0 \int dx~\pi(x) x^m ,~
 r_{\{a_1, \dots, a_m\}} =   r_0 \int dy~\rho(y) y^m,~
t_{\{a_1, \dots, a_m\}} =   t_0 \int dz~\phi(z) z^m $, 
where $q_0$, $r_0$ and $r_0$ are the normalization constants to make
$\pi(x)$, $\rho(y)$ and $\phi(z)$ proper probability distributions
over the interval $[-1,1]$, respectively. Unspecified integrals are
performed over $[-1,1]$. We also assume a similar ansatz for the
conjugate variables. A further complicated assumption about the order
parameter symmetry is generally required in most disordered
systems~\cite{beyond,nishimori}. However, the validity of the RS
ansatz in the current system is strongly supported by a recent report
on the absence of the replica symmetry breaking in gauged systems
where Nishimori's temperature is used. The latter corresponds to 
using the correct priors in decoding~\cite{RSB}, as performed in the 
current analysis.

Under these assumptions, one obtains the free-energy
\begin{eqnarray}
\fl  \calF &=& (1-R_1 -R_2) \ln 2 \nonumber \\
\fl  &-& (1-\alpha - R_1) \left< 
 \ln \left( 1 + \prod_{l=1}^{K_1} x_l \prod_{l^\prime=1}^{K_2} y_{l^\prime}
 \right) \right>_{\pi^{K_1}, \rho^{K_2}} 
- (\alpha - R_2) \left< \ln \left( 1 + \prod_{l=1}^{K_3} z_l \right) 
 \right>_{\phi^{K_3}} \nonumber \\
\fl  &+& (1-\alpha) C_1 \left<  \ln \left( 1 + x \hat{x} \right)
 \right>_{\pi, \hat{\pi}}
 + \alpha C_2 \left< \ln \left( 1 + y \hat{y} \right) 
 \right>_{\rho, \hat{\rho}} 
 +  \alpha C_3 \left< \ln \left( 1 + z \hat{z} \right)
 \right>_{\phi,\hat{\phi}} \nonumber \\
\fl &+& (1-\alpha)  
 \left< \ln \left[ \Tr_s e^{s \xi^{\mbox{\footnotesize\tt u}} F} \prod_{l=1}^{C_1} (1 + s \hat{x}_l )
 \right] \right>_{\xi, \hat{\pi}^{C_1}} \nonumber \\
\fl &+& \alpha \left< 
 \ln \left[ \Tr_{\sigma} e^{\sigma \xi^{\mbox{\footnotesize\tt d}} F} 
 \prod_{l=1}^{C_2} (1+ \sigma \hat{y}_l )  
 \prod_{l^\prime=1}^{C_3} (1+ \sigma \hat{z}_{l^\prime} ) \right]  
 \right>_{\xi, \hat{\rho}^{C_2}, \hat{\phi}^{C_3}}
\label{eq:free-energy}
\end{eqnarray}
where $\langle \cdots \rangle_{P^K}$ denotes an integral of the form 
$\int \prod_{k=1}^Kdx_k P(x_k)(\cdots)$ and 
$\langle f(\xi) \rangle_\xi = (1-p) f(+1) + p f(-1)$.

Varying Eq.(\ref{eq:free-energy}), one obtains a set of saddle-point equations,
\begin{eqnarray}
\fl \left\{
  \begin{array}{l}
   \pi(x) = \left< \delta \left( x - \tanh \left[
   \sum_{l=1}^{C_1-1} \tanh^{-1} \hat{x}_l + \xi^{\mbox{\tt u}} F \right] \right) 
   \right>_{\xi, \hat{\pi}^{C_1-1}} \\
   \rho (y) = \left< \delta \left( y - 
   \tanh \left[ \sum_{l=1}^{C_2-1} \tanh^{-1} \hat{y}_l
   + \sum_{l^\prime=1}^{C_3} \tanh^{-1} \hat{z}_{l^\prime}
   + \xi^{\mbox{\tt d}} F \right] \right)
   \right>_{\xi, \hat{\rho}^{C_2-1}, \hat{\phi}^{C_3}} \\
   \phi (z) = \left< \delta \left( z - 
   \tanh \left[ \sum_{l=1}^{C_2} \tanh^{-1} \hat{y}_l  
    + \sum_{l^\prime=1}^{C_3-1} \tanh^{-1} \hat{z}_{l^\prime} + 
    \xi^{\mbox{\tt d}} F \right] 
   \right) \right>_{\xi, \hat{\rho}^{C_2}, \hat{\phi}^{C_3-1}}
  \end{array}
        \right. \nonumber \\
\fl \left\{
  \begin{array}{l}
   \hat{\pi} (x) = \left< \delta \left( \hat{x} - 
  \prod_{l=1}^{K_1-1} x_l \prod_{l^\prime=1}^{K_2} y_{l^\prime} \right)
  \right>_{\pi^{K_1-1},\rho^{K_2}}  \\
  \hat{\rho} (y) = \left< \delta \left( \hat{y} - 
  \prod_{l=1}^{K_1} x_l \prod_{l^\prime=1}^{K_2-1} y_{l^\prime} \right)
  \right>_{\pi^{K_1},\rho^{K_2-1}}  \\
  \hat{\phi}(z) = \left< \delta \left( \hat{z} - 
  \prod_{l=1}^{K_3-1} z_l \right) \right>_{\phi^{K_3-1}} \\
  \end{array}
\right.
\label{eq:saddle}
\end{eqnarray}

The overlaps 
$\Mup=\frac{1}{(1-\alpha))N} \sum_i \hat{s_i}
\xi_i^{\mbox{\tt u}}$ 
and 
$\Mlo=\frac{1}{\alpha N} \sum_j \hat{\sigma_j} \xi_j^{\mbox{\tt d}}$
serve as performance measures for the error-correcting ability. 
After solving the saddle-point equations (\ref{eq:saddle}), 
these can be calculated as 
\begin{equation}
 \Mup = \int dh~ h_{\mbox{eff}}^{\mbox{\tt u}} (h)~ \sign (h),~ 
  \Mlo = \int dh~ h_{\mbox{eff}}^{\mbox{\tt d}} (h)~ \sign (h), 
\end{equation}
where distributions of effective 
fields $h_{\mbox{eff}}(h)$ are evaluated as 
\begin{eqnarray}
 h_{\mbox{eff}}^{\mbox{\tt u}} (h) &=&  \left< \delta \left( h - 
 \tanh \left[ \sum_{l=1}^{C_1} \tanh^{-1} \hat{x}_l 
 + \xi F \right] \right)  \right>_{\xi, \hat{\pi}^{C_1}}
\nonumber \\
 h_{\mbox{eff}}^{\mbox{\tt d}} (h) &=&  \left< \delta \left(  h - 
 \tanh \left[ \sum_{l=1}^{C_2} \tanh^{-1} \hat{y}_l 
 + \sum_{l^\prime=1}^{C_3} \tanh^{-1} \hat{z}_l 
 + \xi F \right] \right) 
 \right>_{\xi, \hat{\rho}^{C_2},\hat{\phi}^{C_3}}. 
\end{eqnarray}

\subsection{Microscopic analysis -- practical decoding}
As already mentioned, it is computationally hard to perform MPM
decoding (\ref{eq:MPM}) exactly. Instead, the belief propagation (BP)
algorithm~\cite{BP} is widely used for a practical decoding in LDPC
codes. Belief propagation has recently been shown to be equivalent to
the Thouless-Anderson-Palmer (TAP) approach in spin glass
theory~\cite{TAP,BPvsTAP}. We also make use of follow this decoding
algorithm in the current framework.

The BP/TAP approach offers an iterative algorithm to approximately
evaluate marginal posterior distributions based on local dependencies
between syndrome and variables. These local dependencies can be
uniquely identified with conditional probabilities. In the current
system, these become: $q_{\mu l}^n=P(n_l=n|\{\bJ \backslash J_\mu\})$
and $\hat{q}_{\mu l}^n \propto P(J_\mu |n_l=n, \{\bJ \backslash
J_\mu\})$ where $n_l$ and $J_\mu$ represent components of spin
variables $\bs$, $\bsigma$ and syndrome $\bJ$, respectively; $\{\bJ
\backslash J_\mu\}$ denotes the set of syndrome bits excluding
$\mu$-th component. As most syndrome and spin variables are not
directly related, we assign the conditional probabilities only to
pairs $\mu l$ that have non-zero elements in the parity check matrix
$A$.

Evaluating the two types of conditional probabilities using directly
connected components, the BP/TAP algorithm can be generally expressed
as
\begin{eqnarray}
\left \{
\begin{array}{lll}
q_{\mu l}^n&=&\alpha_{\mu l} e^{Fn}
\prod_{\nu \in {\cal M}(l) \backslash \mu}\hat{q}_{\nu l}^n, \cr
\hat{q}_{\mu l}^n&=&\hat{\alpha}_{\mu l} 
\sum_{n_{j \in {\cal L}(\mu) \backslash l}}
\delta(J_\mu;\prod_{j \in {\cal L}(\mu)} n_j)
\prod_{j \in {\cal L}(\mu) \backslash l} q_{\mu j}^{n_j}, 
\end{array}
\right .
\label{eq:general_BP}
\end{eqnarray}
where ${\cal M}(l)$ and ${\cal L}(\mu) $ denote the sets of syndrome
and spin variable indices that are directly linked to spin and
syndrome indices $l$ and $\mu$, respectively; ${\cal M}(l) \backslash
\mu$ represents the set of indices $\nu \in {\cal M}(l) $ excluding
$\mu$ and similarly for ${\cal L}(\mu) \backslash l$ and other sets.
Normalization constants, $\alpha_{\mu l}$ and $\hat{\alpha}_{\mu l}$,
are introduced to make $q_{\mu l}^{n}$ and $\hat{q}_{\mu l}^{n}$
probability distributions of spin variable $n$. A field $F$ is
introduced to represent the prior probability.

Since spin variable $n$ takes only two values $\pm 1$, it is
convenient to express the BP/TAP algorithm using spin averages
$\sum_{n=\pm 1} n q_{\mu l}^{n}$ and $\sum_{n=\pm 1} n \hat{q}_{\mu
  l}^{n}$ rather than the distributions $q_{\mu l}^{n}$ and
$\hat{q}_{\mu l}^{n}$ themselves. As the parity check matrix $A$ is
structured, it may be useful to assign different notation to the
spin averages according to the submatrix to which the pair of indices
$\mu l$ belongs to. We use $x_{\mu l},y_{\mu l}$ and $z_{\mu }$ to
denote $\sum_{n=\pm 1 } n q_{\mu l}^{n}$ when the pair of indices $\mu
l$ belongs to $A_1$, $A_2$ and $A_3$, respectively. Similar notations
$\hat{x}_{\mu l}$,$\hat{y}_{\mu l}$ and $\hat{z}_{\mu l}$ are used for
$\sum_{n=\pm 1} n \hat{q}_{\mu l}^{n}$. Then, the BP/TAP algorithm
(\ref{eq:general_BP}), which is expressed as a set of functional
equations, is reduced to a couple of nonlinear equations
\begin{eqnarray}
\left\{
\begin{array}{l}
x_{\mu l} = \tanh [ \sum_{\nu \in A_1^{\mbox{col}}(l)
\backslash \mu)} \tanh^{-1} 
\hat{x}_{\nu l} + F ],  \\ 
y_{\mu l} = \tanh [ \sum_{\nu \in A_2^{\mbox{col}}(l)
\backslash \mu} \tanh^{-1}
 \hat{y}_{\nu l } 
+ \sum_{\nu \in A_3^{\mbox{col}}(l)} \tanh^{-1} \hat{z}_{\nu l} +  F ], \\
z_{\mu l} = \tanh [ \sum_{\nu \in A_2^{\mbox{col}}(l)} \tanh^{-1} 
\hat{y}_{\nu l}
 + \sum_{\nu \in A_3^{\mbox{col}}(l)
\backslash \mu} \tanh^{-1} \hat{z}_{\nu l} +  F ]. 
\end{array} \right. 
\label{eq:TAP1}
\end{eqnarray}
\begin{eqnarray}
\left\{
\begin{array}{l}
\hat{x}_{\mu l} = \sign (J_\mu) \prod_{
i \in A_1^{\mbox{row}}(\mu) \backslash l}
 x_{\mu l^\prime } \prod_{j \in A_2^{\mbox{row}(\mu)}} y_{ \mu j},  \\
 \hat{y}_{\mu l} = \sign(J_\mu) \prod_{i \in A_1^{\mbox{row}}(\mu)}
 x_{i \mu} \prod_{j \in A_2^{\mbox{row}(\mu)} \backslash l}
 y_{j \mu},  \\
\hat{z}_{\mu l} = \sign(J_\mu) \prod_{j \in A_3^{\mbox{row}}(\mu) \backslash l}
 z_{j \mu}, 
\end{array} \right. 
\label{eq:TAP2}
\end{eqnarray}
where $A^{\mbox{row}}(\mu)$ and $A^{\mbox{col}}(l)$ denote the sets of
non-zero elements in the $\mu$-th row and $i$-th column of matrix $A$,
respectively.

Eqs.(\ref{eq:TAP1}) and (\ref{eq:TAP2}) can be solved iteratively from
appropriate initial conditions (prior means are usually chosen as
initial states). Less then $50$ iterations are typically sufficient
for convergence. After obtaining the solutions, approximated posterior
means can be calculated
\begin{eqnarray}
\langle s_i \rangle &=& \tanh [ \sum_{\nu \in A_1^{\mbox{col}}(i)}
 \tanh^{-1} \hat{x}_{\nu i} + F ] \nonumber \\ 
\langle \sigma_j \rangle &=& 
 \tanh [ \sum_{\nu \in A_2^{\mbox{col}}(j)}  \tanh^{-1} \hat{y}_{\nu i} 
+ \sum_{\nu \in A_3^{\mbox{col}}(j)} \tanh^{-1} \hat{z}_{\nu i} +  
F ], 
\end{eqnarray}
which provides the MPM estimators $\hat{s}_i=\sign(\langle s_i
\rangle)$ and $\hat{\sigma}_j=\sign(\langle \sigma_j \rangle)$.

It can be shown that the BP/TAP framework provides an exact result
when the global structure of the connectivities is graphically
expressed by a tree~\cite{BP}. Unfortunately, it is still unclear how
good are the approximations obtained when a given system does
not admit a tree architecture.

The graphical architecture of LDPC codes generally has many loops,
which implies the BP/TAP framework does not necessarily offer a good
approximation. However, it is conjectured, and partially confirmed,
that a nearly exact result can be obtained, as long as no other
locally stable solutions exists, when the parity check matrix $A$ is
randomly constructed and in the limit $N \to \infty$; this is due to
the fact that the typical loop length scales as $O(\ln N)$ for
randomly constructed matrices, which implies that LDPC codes can be
locally treated as trees ignoring the effect of loops~\cite{Renato}.

\section{Results}
In order to theoretically examine the typical performance that can be
obtained by the linearly combined coding scheme, we solved the saddle
point equations~(\ref{eq:saddle}). Since solving the equations
analytically is generally difficult, we mainly resorted to numerical
methods. The solutions were obtained by iterating the saddle point
equations~(\ref{eq:saddle}), and approximating the distributions by
$O(10^4)$ sample vectors. Less then $50$ iterations were typically
sufficient for obtaining a solution.

Solving the equations for several parameter sets, assuming $\alpha >
R_2/(R_1+R_2)$, we found that the solutions can be classified into
three categories depending on whether overlaps $\Mup$ and $\Mlo$ are
$1$ or not. The first one is referred to the {\em ferromagnetic} (F)
solution ($\Mup = \Mlo = 1$) corresponding to perfect retrieval for
both messages $\calW_1$ and $\calW_2$. The {\em half-ferromagnetic}
(HF) solution which is characterized by $\Mup \neq 1$ and $\Mlo = 1$
implies that only the second message $\calW_2$ is perfectly retrieved,
while $\calW_1$ is not. The last category, termed paramagnetic (P)
solution, describes a decoding failure for both messages being
characterized by $\Mup \neq 1, \Mlo \neq 1$. The ferromagnetic
solution always exits and is locally stable for $C_1 \ge 3$ and $C_3
\ge 3$, while one can find other solutions only for relatively higher
noise levels. As the noise level increases, HF and P solutions emerge
in this order.

The solution that has the lowest free energy among the three becomes
thermodynamically dominant. As the noise level $p$ becomes higher (or
the field $F$ becomes weaker), the dominant state changes from F to HF
and P in this order. Since receivers are required to retrieve only
their own messages, the transition point between HF and P corresponds
to the maximum noise level for error free communication in the second
channel while maximum noise level for the first channel is given by
the transition point between F and HF.

However, this does not imply a successful decoding up to the critical
points in {\em practical} time scales. Practical perfect decoding by
the BP/TAP algorithm is possible only when no suboptimal solutions
exist, which means that the practically achievable limit is given by
the {\em spinodal points} of the HF and P solutions for the first and the
second channels respectively; i.e., the point where new suboptimal
solutions emerge. A similar phenomena has been reported before for
similar systems~\cite{KMS_PRL,KMSV}.

\begin{figure}
 \psfrag{p_1}{$p_1$}
 \psfrag{p_2}{$p_2$}
  \begin{center}
  \includegraphics[width=9cm,clip]{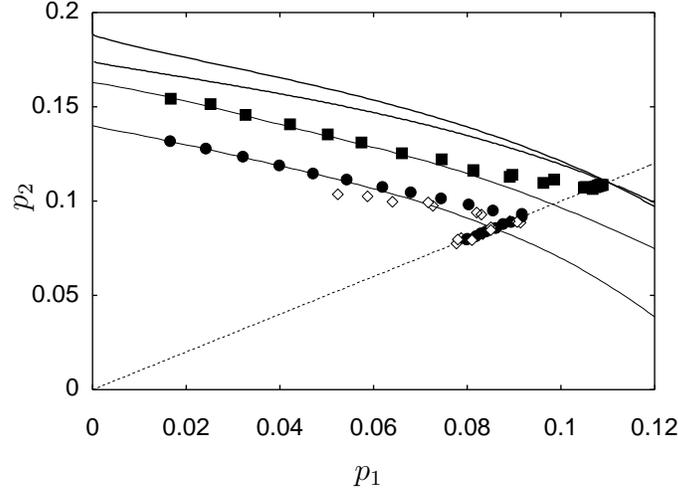}
  \end{center}
  \caption{Optimal and practical performance of the MPM decoder
    calculated by methods of statistical mechanics for different
    $\alpha$ values. For the first channel, the optimal performance
    is given by the thermodynamical transition between F and HF
    solutions while the transition between HF and P solutions marks
    the optimal performance for the second channel. On the other
    hand, the practical performance is given by the spinodal points of
    the HF and P solutions for the first and the second channels,
    respectively. Monte Carlo solutions based on $10^4$ sample vectors
    were employed for solving the saddle-point equation
    (\ref{eq:saddle}). The standard deviation values resulting from
    10 trials are smaller than the symbol size.
    
    The black squares and the black circles denote the optimal and the
    practical performances for the linearly combined coding scheme,
    where code parameters are set to $C_1 = C_3 = 3, C_2=4$,
    $R_1=R_2=1/4$. Diamond symbols denote the maximum noise levels
    for decoding success by the BP/TAP algorithm, determined from 50
    experiments. The error bars are smaller than the symbols.
    
    Broken lines denote the optimal and practical performances of the
    timesharing for corresponding LDPC codes. The two lines in the
    upper right are Cover's and timesharing capacities calculated in
    the information theory. }
 \label{fig:phase_diagram}
 \end{figure}

Fig.\ref{fig:phase_diagram} shows the maximum noise levels for perfect
decoding of the linearly combined coding method obtained for $C_2=4$
and $0$ fixing $C_1=C_3=3$; $C_2=0$ corresponds to the sharing scheme
for which $A_2=0$. One can find that both optimal and practical
performances of the MPM decoder are improved by the introduction of
the additional submatrix $A_2$, as anticipated, in spite of the fact
that the parameter $C_2(=4)$ is not optimally tuned. This result may
induce the hope that Cover's limit can be saturated by optimally
tuning the submatrices.

 However, our analysis contradicts this conjecture. Solving
 Eq.(\ref{eq:saddle}) in the limit $C_3 \to \infty$ and $C_1$ or $C_2
 \to \infty$ is feasible; it is known that the MPM decoder provides
 the optimal performance in this limit while practical BP/TAP decoding
 becomes difficult. The three solutions correspond to those 
 already mentioned before, but can be analytically expressed as:
 \begin{itemize}
  \item {\em F solution:} Both messages are decodable($\Mup = \Mlo = 1$).
        The corresponding solutions and free energy are
        \begin{equation}
         \left\{
          \begin{array}{l}
           \pi (x) = \delta (x-1) \\
           \rho (y) = \delta (y-1) \\
           \phi (z) = \delta (z-1),
          \end{array}
          \right.
         ~~\left\{
            \begin{array}{l}
             \hat{\pi} (\hat{x}) = \delta(\hat{x}-1) \\
           \hat{\rho} (\hat{y}) = \delta(\hat{y}-1) \\
           \hat{\phi} (\hat{z}) = \delta(\hat{z}-1), 
            \end{array}
          \right.
        \end{equation}
        \begin{equation*}
         \calF = -(1-2p) F.
        \end{equation*}
        
  \item {\em HF solution:} Message $\calW_2$ is 
        only decodable($\Mup \neq 1, \Mlo = 1$).
        \begin{equation}
         \left\{
          \begin{array}{l}
           \pi(x) = \langle \delta(x - \tanh \xi F) \rangle_\xi \\
           \rho(y) = \delta(y-1) \\
           \phi(z) = \delta(z-1).
          \end{array}
          \right.
          ~~\left\{
             \begin{array}{l}
              \hat{\pi}(\hat{x}) = \delta(\hat{x}).
             \hat{\rho}(\hat{y}) = \delta(\hat{y}) \\
              \hat{\phi}(\hat{z}) = \delta(\hat{z}-1),
             \end{array}
          \right.
        \end{equation}
        \begin{equation*}
         \calF = (1-\alpha-R_1) \ln 2 -(1-2p) F + 
          (1-\alpha) [p \ln p + (1-p) \ln (1-p) ].
        \end{equation*}
        
  \item {\em P solution:} 
        Both messages are not decodable($\Mup \neq 1, \Mlo \neq 1$).
        \begin{equation}
         \left\{
          \begin{array}{l}
           \phi(z) = \langle \delta(z- \tanh \xi F) \rangle_\xi \\,
            \rho(y) = \langle \delta(y- \tanh \xi F) \rangle_\xi \\
           \pi(x) = \langle \delta(x - \tanh \xi F) \rangle_\xi, 
          \end{array}
    \right.
    ~~\left\{
       \begin{array}{l}
        \hat{\phi}(\hat{z}) = \delta (\hat{z})  \\
        \hat{\rho}(\hat{y}) = \delta(\hat{y})  \\
        \hat{\pi}(\hat{x}) = \delta(\hat{x}),
       \end{array}
     \right.
        \end{equation}
        \begin{equation*}
         \calF = (1-R_1-R_2) \ln 2 - (1-2p)F + p \ln p + (1-p) \ln (1-p).
        \end{equation*}
 \end{itemize}
 
 Examining the critical condition for decoding success in each
 channel, and comparing the free energy of the solutions, one obtains
 the capacity region of the linearly combined coding scheme
 \begin{equation}
  \left\{
   \begin{array}{l}
    R_2 < \alpha [ 1 - H(p_2) ] \\
    R_1 < (1-\alpha) [ 1 - H(p_1) ]. 
   \end{array}
   \right.
 \end{equation}
 This is, unfortunately, identical to the timesharing capacity which
 can be achieved by a simple concatenation of two independent codes.
 This result implies that the advantage of the linearly combined
 coding scheme vanishes as the submatrices become dense and this
 method cannot saturate Cover's limit.


\section{Summary and Conclusion}
In this paper, we have examined the performance of linearly combined
LDPC codes, for information transmission in a broadcast channel. Our
analysis shows that the capacity of the suggested coding scheme is
upperbounded by the timesharing capacity, in spite of the apparent
improvement in both optimal and practical performance with respect to
LDPC based timesharing codes characterized by finite connectivity
values.

The reason for the failure of linearly combined LDPC codes to saturate
Cover's limit may be explained by the codeword structure produced by
this scheme. In his proof, Cover optimized the code performance by
introducing a specific structure termed the {\it cloud coding},
employing an auxiliary random variable $\calU$ as in
Eq.(\ref{eq:capacity}). In cloud coding, a codeword $\calX$ is
randomly generated according to $P(\calX|\calU)$ around a {\em cloud
center} $\calU$ sampled from $P(\calU)$. Knowing this structure, one
can use the cloud center $\calU$ and the coset $\calX_c = \calX -
\calU$ for encoding $\calW_2$ and $\calW_1$, respectively.
   
In the case of binary symmetric channels, the optimal cloud center
$\calU$ can be obtained by sampling $N$ bit unbiased vectors for
which the entropy per bit can be maximized to $1$. On the other hand,
one can produce the optimal coset $\calX_c$ by independently and
randomly generating each bit using a uniform bias $0<\delta<1$, which
provides an entropy $H_2(\delta)$ per bit.

In an ideal situation, a noise vector $\bxi_1$ which is biased with a
flip probability $p_1$ is added to the coset $\calX_c$ in the first
channel. This implies that the entropy of the received coset becomes
$H_2(\delta * p_1)$ per bit while the entropy of the noise vector is
$H_2(p_1)$ per bit.  Since one can use the difference between the
entropies to convey the information of $\calW_1$, the capacity of the
first channel becomes $R_1 < H_2(\delta * p_1)-H_2(p_1)$, which is the
second inequality of Eq.(\ref{eq:capacity}).  On the other hand, for
the second channel, characterized by a flip rate $p_2$, the coset
$\calX_c$ together with a channel noise $\bxi_2$ serves as a single
noise vector for which the entropy becomes $H_2(\delta* p_2)$ per bit.
As the entropy of the received cloud center can be maximized to $1$
per bit, this means that the capacity of the second channel is given
by $R_2 < 1-H_2(\delta*p_2)$, which is the first inequality of
Eq.(\ref{eq:capacity}).

In linearly combined coding scheme $({G_2^T \atop G_3^T}) \calW_2 +
({G_1^T \atop 0}) \calW_1$, $({G_2^T \atop G_3^T}) \calW_2 $ becomes
almost random, which may serve as the optimal cloud center. However,
the second part $({G_1^T \atop 0}) \calW_1$, that corresponds to the
coset, is somewhat structured, differing from the optimal choice of
uniformly biased random vectors.

In order to compare the structured coset with the optimal one, let us
fix the maximum entropy per bit of $({G_1^T \atop 0}) \calW_1$, which
equals $1-\alpha$, to that of the optimal coset $H_2(\delta)$. Then,
one can show that the entropy of the corrupted coset with flip
probability $p$ per bit always increases from $H_2(p * \delta)$ to
$(1-\alpha)+\alpha H_2(p)= H_2(\delta)+H_2(p) \ge H_2(p * \delta)$.
This means that the critical rate of the first channel increases from
$H_2(\delta * p_1)-H_2(p_1)$ to $(1-\alpha)\left [1-H_2(p_1) \right ]$
while that of the second channel reduces from $1-H_2(p_2)$ to $\alpha
\left [1-H_2(p_2) \right ]$. This trade-off between the capacities of
the two channels limits the performance of linearly combined coding
scheme to the timesharing limit, that is always within Cover's
capacity region.

In conclusion, while the suggested linearly combined LDPC coding scheme
provides an improved performance over LDPC based timesharing codes for
finite connectivity constructions, in both theoretical and practical 
limits, it cannot go beyond the theoretical timesharing limit; for 
that to happen, different coding schemes should be examined.

\vspace*{.5cm}

\vspace{3mm}

\ack
Support by Grants-in-aid, MEXT (13680400 and 13780208) and JSPS (YK),
The Royal Society and EPSRC-GR/N00562 (DS) is acknowledged.

\end{document}